\documentclass{aastex631}
\makeatletter

\newcommand{\Rmnum}[1]{\expandafter\@slowromancap\romannumeral #1@}
\makeatother

\shorttitle{Secondary flux emergence in ephemeral region}
\shortauthors{Yang et al.}

\graphicspath{{./}{figures/}} 
\begin{document}

\title{Complexity of emerging magnetic flux during lifetime of solar ephemeral regions}

\author{Hanlin Yang}
\affiliation{National Astronomical Observatories, Chinese Academy of Science, Beijing 100101, People’s Republic of China}
\affiliation{University of Chinese Academy of Sciences, 100049, Beijing, People’s Republic of China}
\affiliation{State Key Laboratory of Solar Activity and Space Weather, Beijing 100190, people's Republic of China}

\author{Chunlan Jin}
\affiliation{National Astronomical Observatories, Chinese Academy of Science, Beijing 100101, People’s Republic of China}
\affiliation{State Key Laboratory of Solar Activity and Space Weather, Beijing 100190, people's Republic of China}

\author{Zifan Wang}
\affiliation{National Astronomical Observatories, Chinese Academy of Science, Beijing 100101, People’s Republic of China}
\affiliation{State Key Laboratory of Solar Activity and Space Weather, Beijing 100190, people's Republic of China}

\author{Jingxiu Wang}
\affiliation{National Astronomical Observatories, Chinese Academy of Science, Beijing 100101, People’s Republic of China}
\affiliation{University of Chinese Academy of Sciences, 100049, Beijing, People’s Republic of China}
\affiliation{State Key Laboratory of Solar Activity and Space Weather, Beijing 100190, people's Republic of China}

\begin{abstract}

As a relatively active region, ephemeral region (ER) exhibits highly complex pattern of magnetic flux emergence. We aim to study detailed secondary flux emergences (SFEs) which we define as bipoles that they appear close to ERs and finally coalesce with ERs after a period. We study the SFEs during the whole process from emergence to decay of 5 ERs observed by the Helioseismic and Magnetic Imager (HMI) aboard Solar Dynamics Observatory (SDO) . The maximum unsigned magnetic flux for each ER is around $10^{20}$ Mx. Each ER has tens of SFEs with an average emerging magnetic flux of approximately 5$\times10^{18}$ Mx. The frequency of normalized magnetic flux for all the SFEs follows a power law distribution with an index of -2.08 . The majority of SFEs occur between the positive and negative polarities of ER , and their growth time is concentrated within one hour. The magnetic axis of SFE is found to exhibit a random distribution in the 5 ERs. We suggest that the relationship between SFEs and ERs can be understood by regarding the photospheric magnetic field observations as cross-sections of an emerging magnetic structure. Tracking the ERs' evolution, we propose that these SFEs in ERs may be sequent emergences from the bundle of flux tube of ERs, and that SFEs are partially emerged $\Omega$-loops.

\end{abstract}


\section{Introduction} \label{sec:1}

 Ephemeral regions (ERs) are small, magnetic bipolar configurations on the surface of the Sun. They are not biased towards or away from active regions, making them observable at any time during a solar cycle at nearly any position \citep{harvey1973ephemeral}. During the emergence phase, their bipoles tend to separate from each other, while in the later stages, they either approach and cancel, or separate and diffuse. The average flux of a typical ER is approximately $10^{20}$ Mx. Many studies have been conducted to observe and statistically analyze the emerging magnetic flux of ERs. But they give different magnetic flux range, such as 1.5$\times10^{19}$ Mx \citep{1988SoPh..116....1W}, 2.5$\times10^{19}$ Mx \citep{martin1990small}, 5.0$\times10^{19}$ Mx \citep{1992ASPC...27..108W} and 1.3$\times10^{19}$ Mx \citep{1997ApJ...487..424S} based on magnetic measurements from different instruments . Using data from the Helioseismic and Magnetic Imager (HMI) aboard Solar Dynamics Observatory (SDO), \citet{yang2013properties} find that during the emergence process, ERs display three kinds of kinematic performances: separation, shift of the polarities' center, and rotation of the ER's axis. The average separation velocity, shift velocity and angular speed are 1.1 km ${\rm s}^{-1}$, 0.9 km ${\rm s}^{-1}$, and $0.6^{\circ}$ ${\rm minute}^{-1}$, respectively. 

On a smaller scale, \citet{wang2012solar} identify simultaneous bipolar emergences within the network. They define these bipoles as intranetwork ephemeral regions (IN ERs). Then they find that the smallest IN ERs have a maximum unsigned magnetic flux of several $10^{16}$ Mx. Interestingly, a few IN ERs show repeated shrinkage-growth or growth-shrinkage in magnetic flux , similar to magnetic floats in the dynamic photosphere, which is an $\Omega$-loop moving up and down while slightly rotating \citep{2012SoPh..278..299W}. The smallest ER appears to consist of a single bipole, although it has been reported that larger-scale ones are composed of a series of bipoles that emerge next to the original ones simultaneously and persistently \citep{martin1990elementary}.

Inspired by Martin's concept \citep{martin1990elementary}, we have analyzed five typical ERs observed by SDO/HMI and discovered that the bipolar configurations of ERs do not evolve regularly by growing individually from small to large and eventually disappearing. Each region consists of tens of bipolar emergences, making the phenomenon more complex than previously thought. We define the regions that occupy the vast majority of magnetic flux and exhibit slow and visually discernible topological changes as main polarities of ER, and define these small bipoles that appear close to the main polarities and finally coalesce with the main polarities as secondary flux emergences (SFEs). In this article, we attempt to address several questions concerning ERs: the number of SFEs in a ER, the SFEs' flux contribution to the ER magnetic flux, the association between SFEs and ER, the mechanisms of SFEs generation, and SFEs' roles in triggering small-scale solar activity. We propose a potential sub-photospheric magnetic field configuration of generally $\Omega$-shaped loops by analyzing magnetograms throughout the evolving ERs. In the discussion section, we briefly mention a small-scale eruptive-activity event and its atmospheric response solely induced by SFEs.

Here, we outline the upcoming sections of the paper. In Section \ref{sec:2}, we detail the observations and data analysis methods. In Section \ref{sec:3} we present the statistical results. We summarize and discuss our findings in Section \ref{sec:4}.

\section{observations and data analysis} \label{sec:2}

\begin{figure}[h]
\centering
\includegraphics[width=0.6\textwidth]{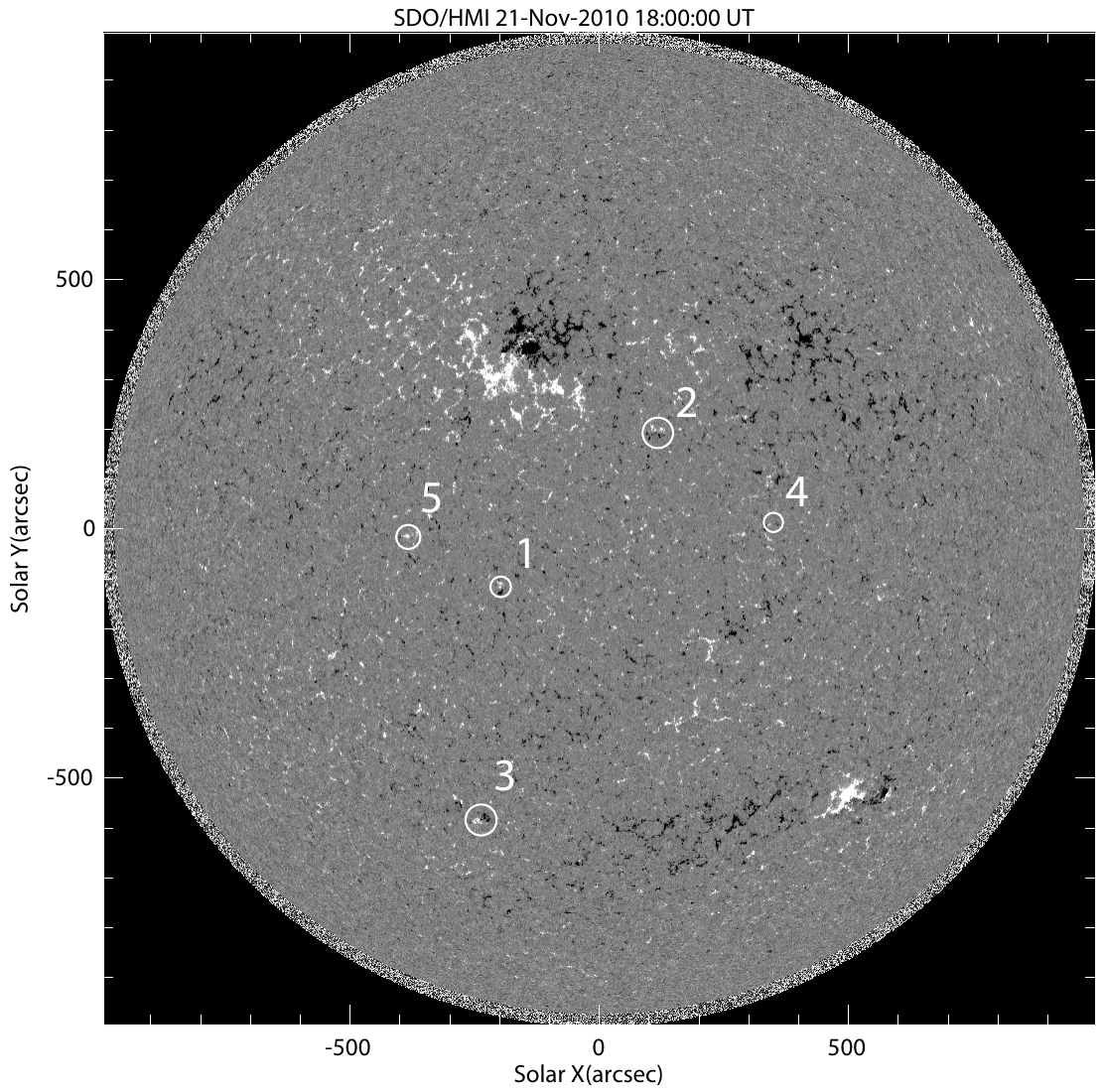}
\caption{Positions of five selected ERs. ER1 evolved from 21-Nov 10:00 UT to 24-Nov 07:00 UT, ER2 from 20-Nov 17:00 UT to 22-Nov 15:00 UT, ER3 from 20-Nov 17:00 UT to 23-Nov 15:00 UT, ER4 from 22-Nov 02:00 UT to 23-Nov 08:00 UT, and ER5 from 20-Nov 13:00 UT to 21-Nov 18:00 UT. The five ERs are situated far from active regions and within 60 degrees of solar disk. Therefore, the projection effect correction is not made. \label{fig:fulldisc}} 
\end{figure}

 SDO/HMI measures line-of-sight magnetic fields in the solar photosphere using the 6173 $\rm{\AA}$ Fe $\rm{\Rmnum{1}}$ absorption line. 
 The pixel size is 0.5 $''$ and the cadence is 45 s. Five ERs are selected based on four days' continuous observations from 2010 November 20 to 24, and these ERs are outlined by white circles in Figure \ref{fig:fulldisc}. The five ERs are far away from existing large-scale magnetic structures and active regions. Furthermore, they are located around the disk center and exhibit a complete evolution process. In this study, all magnetograms are derotated to fixed positions in order to counteract the effect of differential rotation. To improve the signal-to-noise ratio and accurately capture SFEs, we apply a temporal smoothing technique by averaging magnetograms in 4 mins, as well as a spatial smoothing technique through overlay averaging with a 2-pixel radius.  The noise level is estimated to be around 7 G per pixel, which is determined by fitting the magnetic field to a Gaussian distribution (\citet{2001ApJ...555..448H}, \citet{2004SoPh..219...39L}, \citet{2011ApJ...731...37J} and \citet{ 2020ApJ...889L..26J}). In this study, the magnetic threshold of 7 G is adopted. The magnetic flux of the five ERs is obtained by tracking their magnetic evolutions .

\begin{figure}[h]
\centering
\includegraphics[width=0.7\textwidth]{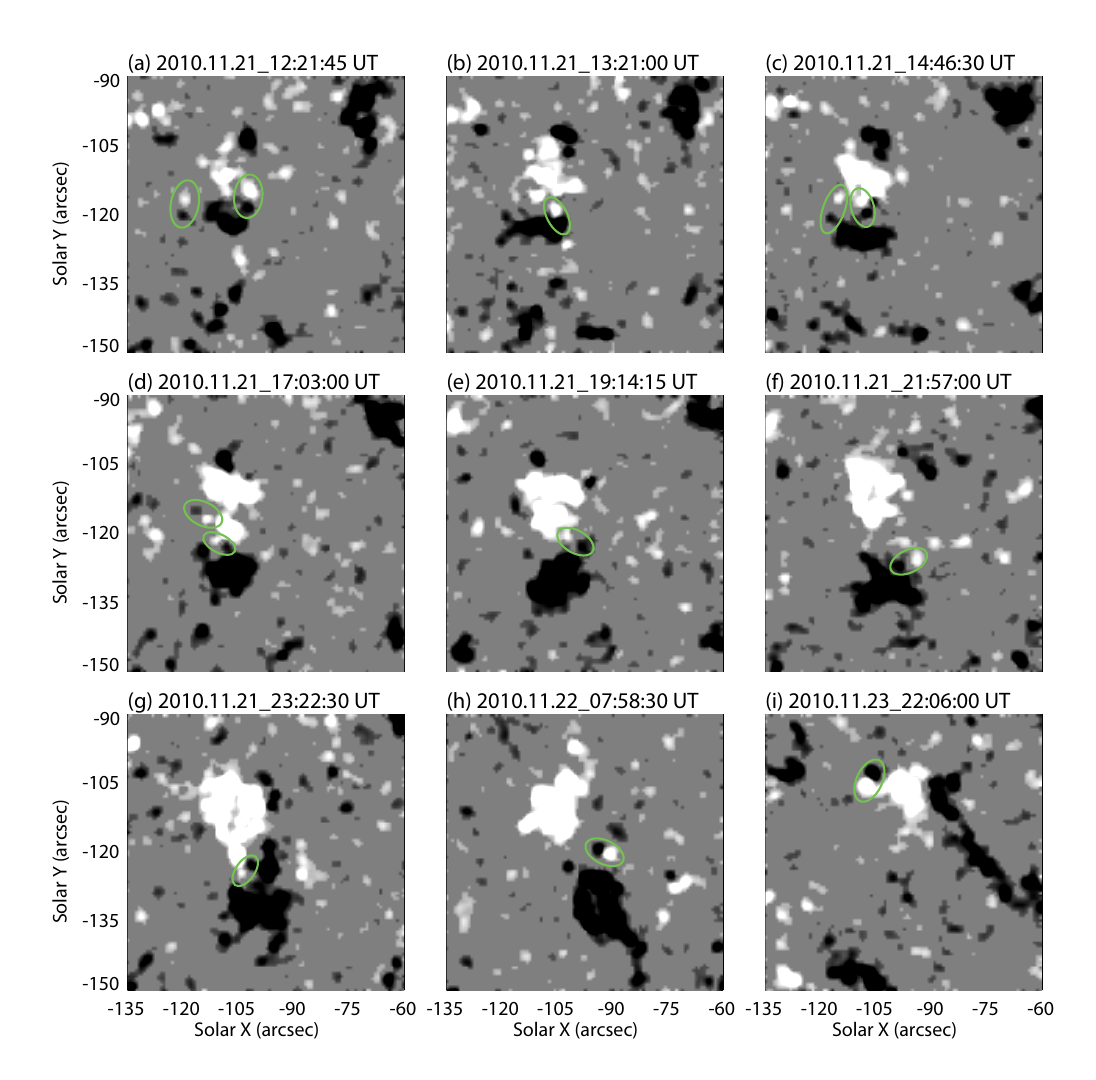}
\caption{Twelve representative instances of SFEs. Magnetograms reveal that SFEs may appear anywhere in the surroundings of the main polarities. As shown in panels (a)-(h), most of the SFEs eventually merge with the main polarities, which lead to the ER's growing in magnetic flux. However, for a small fraction of SFEs, such as the example in panel (i), magnetic cancellation is the ultimate outcome. To enhance visual contrast, we set pixel values below the noise level (7 G) to zero. The magnetograms saturate at 21 G .\label{fig:emergence}} 
\end{figure}

 First , we use visual recognition to identify and track SFEs. As shown in Figure \ref{fig:emergence}, several SFEs are observed in the ER1. SFEs typically manifest as small bipoles that appear around the main polarities. They may occur in various locations in the surroundings of the two main polarities, such as the intermediate region between two main polarities (e.g., panels (b), (d), (e) and (g)), the periphery of a specific main polarity (e.g., panels (f) and (i)), or even in the region that are not adjacent to either main polarity (as indicated by the green circle on the left side of panel (a) in Figure \ref{fig:emergence}). Additionally, the magnetic axes of SFEs are not consistent with that of the host ER. Sometimes, the magnetic axis of the SFE even has opposite direction with that of the ER, such as the SFE shown in panel (h).

\begin{figure}[h]
\centering
\includegraphics[width=1.0\textwidth]{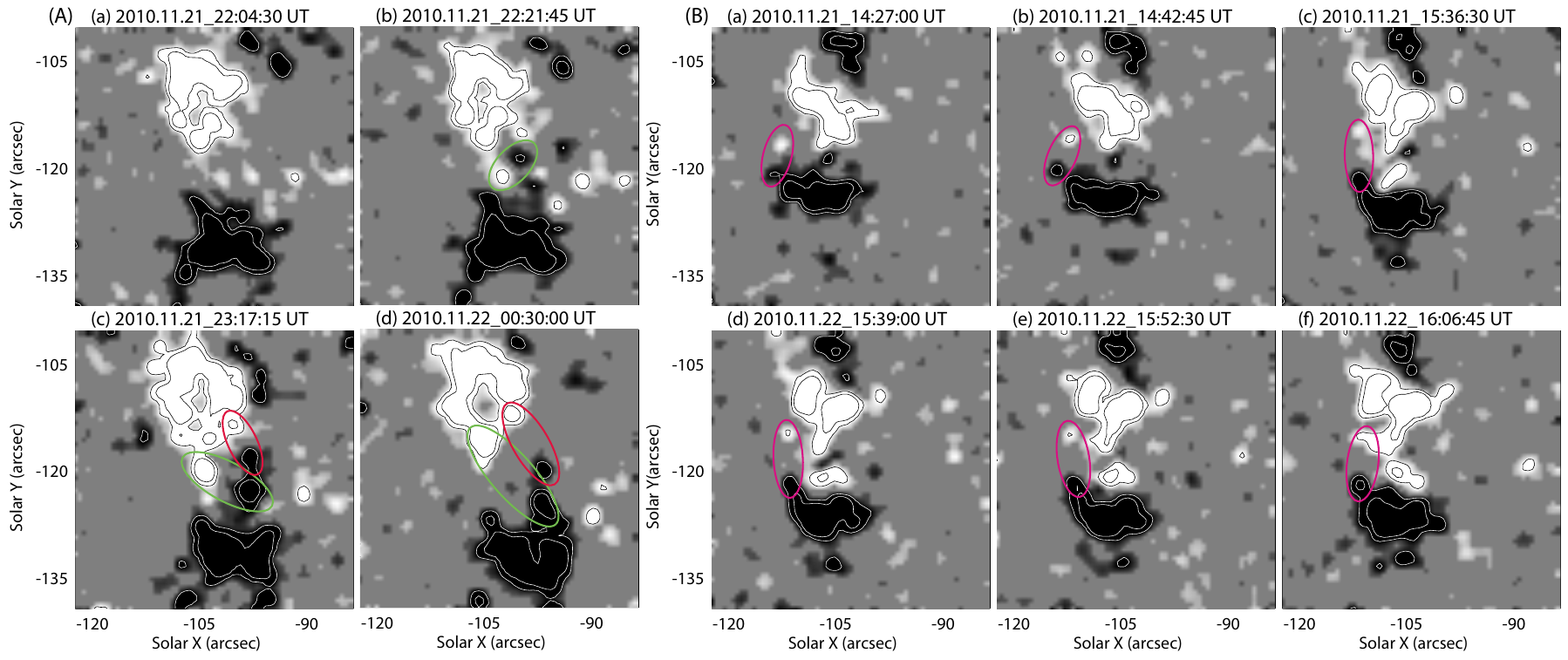}
\caption{Evolution of two SFEs in ER1, with the second one exhibiting magnetic float behavior. Panel (A), subfigure (a) shows the magnetic background before SFE. (b) depicts the developing phase. In subfigure (c), the small positive configuration begins to merge with the main polarity. This moment marks the time when the magnetic flux reaches its highest level and yields the maximum flux. (d) illustrates the state after the complete merging of the positive and negative polarities with the main polarity. We also highlight, with red ellipses in subfigure (c) and (d), another pair of SFE that emerges simultaneously. Panel (B) presents a SFE exhibiting magnetic float behavior. Similarly, we set pixel values below the noise level to zero. \label{fig:contour}} 
\end{figure}

 Secondly , we obtain the magnetic flux of each SFE. Panel (A) in Figure \ref{fig:contour} illustrates the detailed process of identifying a specific SFE. In subfigure (a) of panel (A), no SFE are observed between the two main polarities. In subfigure (b), a pair of small bipolar features, indicated by a green ellipse, is extracted by the threshold of 21 G. Subsequently, at 23:17:15 UT as shown in subfigure (c), and both polarities of SFE reach maximum flux , and its positive polarity, which is marked by the green ellipse, begins to merge with the main positive polarity. Thus, we record the magnetic flux of the SFE at this moment. In subfigure (d), its negative polarity also merges with the main negative polarity. Both polarities of the SFE lose their distinctiveness, which means that they could no longer be individually identified at the resolution of HMI and henceforth evolve together with the main polarities. It is worth noting that subfigures (c) and (d) highlight another pair of SFE indicated by red ellipses, and we select the moment in subfigure (d) when their magnetic flux reaches its maximum.

It should be noted that magnetic floats can also affect the identification of SFEs, as first mentioned by \citet{wang2012solar}, and they occur at the level of the intranetwork. The panel (B) in Figure \ref{fig:contour} displays a SFE sample, which exhibits the properties of a magnetic float. The obvious feature is the pattern of growth-decay-growth in magnetic flux magnitude . In subfigure (a) of panel (B), a pair of bipolar features emerges to the left of the main polarities with a magnetic flux of 1.0 $\times$ $10^{18}$ Mx. Subsequently, this pair of SFE begins to grow, reaching a magnetic flux of 1.5 $\times$ $10^{18}$ Mx in subfigure (b). However, the bipolar features start to diminish, and in subfigure (c), the magnetic flux reduces to only 5.8 $\times$ $10^{17}$ Mx. Afterward, the SFE start to intensify again, with magnetic flux values of 8.2 $\times$ $10^{17}$ Mx and 1.1 $\times$ $10^{18}$ Mx in subfigures (d) and (e), respectively. At the moment depicted in subfigure (f), its positive and negative polarities fully merge with the main polarities, becoming indistinguishable. Sometimes, magnetic floats can annihilate below the magnetic noise, affecting our judgement of whether the bipolar features of the SFE at different time represent two distinct pairs of bipolar features or different manifestations of the same pair of bipolar features. In this study, if the structure keeps same polarity at the same position all the time, we count only once no matter how weak it is in a period of time.

Thirdly, we procure the maximum unsigned magnetic flux of each SFE . 
To determine the range of maximum unsigned magnetic flux, it is essential to identify the moment when bipoles attain their highest level. This situation can occur in two ways. In the first scenario, the magnetic flux reaches its maximum, but neither of the two polarities becomes incorporated into the main bipole. In this case, we delineate the area and calculate the magnetic flux. In the second scenario, at least one magnetic patch is eventually absorbed into the main regions and grows along with them. As the secondary polarities lose their independent characteristics after this point, we select a particular moment before the merging for computation, at which the unsigned magnetic flux reaches its maximum. This mechanism is explained in Figure \ref{fig:contour}. In Section \ref{sec:3}, we will describe the distribution features in detail.

 Furthermore, we also obtain the time for each SFE to grow to its maximum flux. 
For the vast majority of SFEs, we record the initially appearing moment (${\rm {T}}_{\rm{sta}} $) and moment with maximum flux (${\rm {T}}_{\rm {max}}$), so that we calculate the growth time as ${\rm {T}}_{\rm {gro}}$=${\rm {T}}_{\rm {max}}-{\rm {T}}_{\rm {sta}}$ .

Finally, we utilize flux-weighted centers to record the location of positive and negative patches for each SFE to investigate the magnetic axis orientation. 
Additionally, we calculate the distance between the SFE and the main polarities, denoted as $\rm{d_i}$ , which is expressed as the length between the absolute flux-weighted center of the SFE and that of the ER . For comparative analysis, we also recorded the maximum separation distance ($\rm{D}$) between the positive and negative polarities of each ER .

\section{results} \label{sec:3}

\begin{deluxetable*}{ccccccccccc}
\tablenum{1}
\tablecaption{Mean values for 5 ephemeral regions\label{tab:mean value}} 
\tablewidth{0pt}
\tablehead{
\colhead{Ephemeral region} & \colhead{$\rm{\Phi^{ER}_{peak}} $} & \colhead{Duration} & \colhead{Number of} & \colhead{$\rm{\Phi^{SFE}_{i}} $} & \colhead{$\sum\rm{\Phi^{SFE}_i}$} & \colhead{$\rm{\Phi^{SFE}_i}/\rm{\Phi^{ER}_{peak}} $} & \colhead{$\rm{T_{gro}} $} & \colhead{$\rm{d_i}$} & 
\colhead{$\rm{D}$} 
 \\
\colhead{$\#$} & \colhead{($10^{20}$Mx)} & \colhead{(h)} & \colhead{secondary flux emergence} &
\colhead{($10^{18}$Mx)} & \colhead{($10^{20}$Mx)} & & \colhead{(h)} & \colhead{($10^{3}$km)} & \colhead{($10^{3}$km)}
}
\startdata
ER1 & 1.4 & 69.0 & 39 & 0.6-14.6 & 1.2 & 0.5\%-10\% & 0.025-5.3 & 1.1-17.6 & 18 \\
ER2 & 1.6 & 46.0 & 20 & 1.0-15.7 & 1.2 & 0.6\%-10\% & 0.075-2.1 & 2.5-15.7 & 16 \\
ER3 & 2.1 & 70.0 & 41 & 1.0-40.0 & 2.2 & 0.5\%-20\% & 0.1-4.2 & 0.6-19.5 & 16 \\
ER4 & 0.6 & 30.0 & 21 & 0.5-13.0 & 0.8 & 0.8\%-20\% & 0.05-1.5 & 0.5-14.3 & 12 \\
ER5 & 2.1 & 29.0 & 51 & 1.3-26.0 & 2.2 & 0.6\%-10\% & 0.025-7.6 & 1.3-28.8 & 20 \\
\enddata
\tablecomments{Several average features of the studied SFEs. Specifically, $\rm{\Phi^{ER}_{peak }}$ denotes the peak magnetic flux of the corresponding ER, while $\sum{\rm{\Phi^{SFE}_i }}$ indicates the total magnetic flux $\rm{\Phi^{SFE}_i}$ for each of these ERs. Additionally, we report the growth time $\rm{T_{gro}} $ and the distance $\rm{d_i}$ between each SFE and its corresponding ER, as well as the range of magnetic flux $\rm{\Phi^{SFE}_{i}} $ for the SFEs. Finally, we also provide the maximum separation distance $\rm{D}$ of the corresponding ER bipole. }
\end{deluxetable*}

Here we present the statistical features of SFEs.

\emph{Flux distribution}: Table \ref{tab:mean value} presents the basic properties of SFEs for five ERs. The unsigned magnetic flux of the five ERs is approximately $10^{20}$ Mx. The values of $\rm{\Phi^{SFE}_i}$ (magnetic flux of each SFE) range from $10^{17}$ to $10^{19}$ Mx. Furthermore, for each ER, the larger the peak magnetic flux ($\rm{\Phi^{ER}_{peak}} $) of the ER, the stronger the upper limit of individual magnetic flux for SFE ($\rm{\Phi^{SFE}_i}$). We find that the peak magnetic flux of an ER is not positively correlated with the number of SFEs. For example, ER1 has a smaller $\rm{\Phi^{ER}_{peak}} $ compared to ER2, yet it exhibits a higher number of SFEs. This could be attributed to the stronger magnetic field strength of ER2, which counteracts the extent of disruption caused by turbulent convection \citep{2014LRSP...11....3C}.

\begin{figure}[h]
\centering
\includegraphics[width=1.0\textwidth]{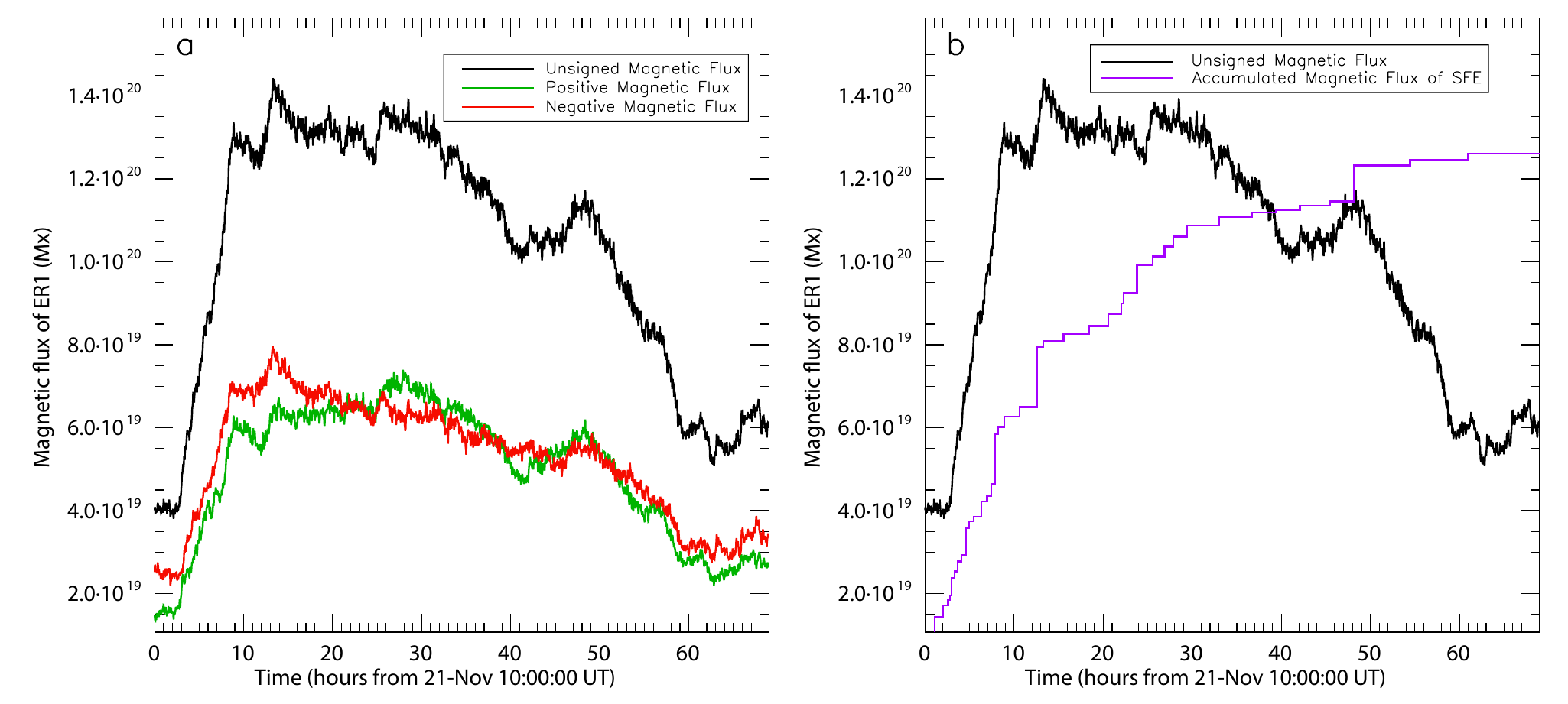}
\caption{Variation of magnetic flux in ER1. The black line represents the unsigned magnetic flux of ER1, with the green line depicting its positive magnetic flux and the red line representing the negative magnetic flux. In panel b, the purple line illustrates the accumulated $\rm{\Phi^{ER}_{peak}} $. \label{fig:co_flux}} 
\end{figure}

 Figure \ref{fig:co_flux} illustrates the flux evolution of ER1 and its accumulated $\rm{\Phi^{SFE}_i}$. The latter is the result of summing the magnetic flux of each SFE when it reaches its respective $\rm{\Phi^{SFE}_i}$. In panel a, the simultaneous variation is observed between positive and negative magnetic flux, though they do not entirely overlap, possibly due to the uneven background magnetic field. Panel b displays the changes in ER magnetic flux and the accumulated $\rm{\Phi^{SFE}_i}$, both exhibiting rapid growth in the first approaching 10 hours. Then, ER maintains the maximum magnetic flux in subsequent about 20 hours. Later, ER flux gradually decreases due to magnetic cancellation \citep{1985AuJPh..38..929M}, yet a small number of SFEs continue to contribute to the ascent of accumulated $\rm{\Phi^{SFE}_i}$. In panel b, we observe that at 50th hour (which corresponds to the late phase of ER), a pronounced SFE occurs, resulting in a noticeable increase in ER flux as well. SFE plays a crucial role throughout the lifetime of ER.

Another intriguing observation is that according to Table \ref{tab:mean value}, we observe that $\sum\rm{\Phi^{SFE}_i}$ is comparable to $\rm{\Phi^{ER}_{peak}} $, and in some cases like ER3, ER4, and ER5, $\sum\rm{\Phi^{SFE}_i}$ exceeds $\rm{\Phi^{ER}_{peak}} $. This implies that seemingly inconspicuous SFEs contribute to the majority of the magnetic flux in ER. In other words, ER is essentially composed of these smaller magnetic flux tubes converging. The situation where $\sum\rm{\Phi^{SFE}_i}$ surpasses $\rm{\Phi^{ER}_{peak}} $ can be attributed to the following three factors. First, SFEs do not all reach their peak strength simultaneously. When we determine the moment of peak strength in ER, some SFEs might not appear or reach their maximum strength. The second point involves that, with the onset of flux cancellation, the magnetic flux in the main polarities keeps diminishing. Consequently, flux cancellation also leads to a decrease in $\rm{\Phi^{ER}_{peak}} $ . Thirdly, during the emergence of magnetic flux tubes, deformations can occur. If split flux tubes emerge on the photosphere in the form of twisted structures, it essentially leads to repeated statistical counting. This is also responsible for the higher value of $\sum\rm{\Phi^{SFE}_i}$. We will elaborate on this aspect in Section \ref{sec:4}.

\begin{figure}[h]
\centering
\includegraphics[width=0.8\textwidth]{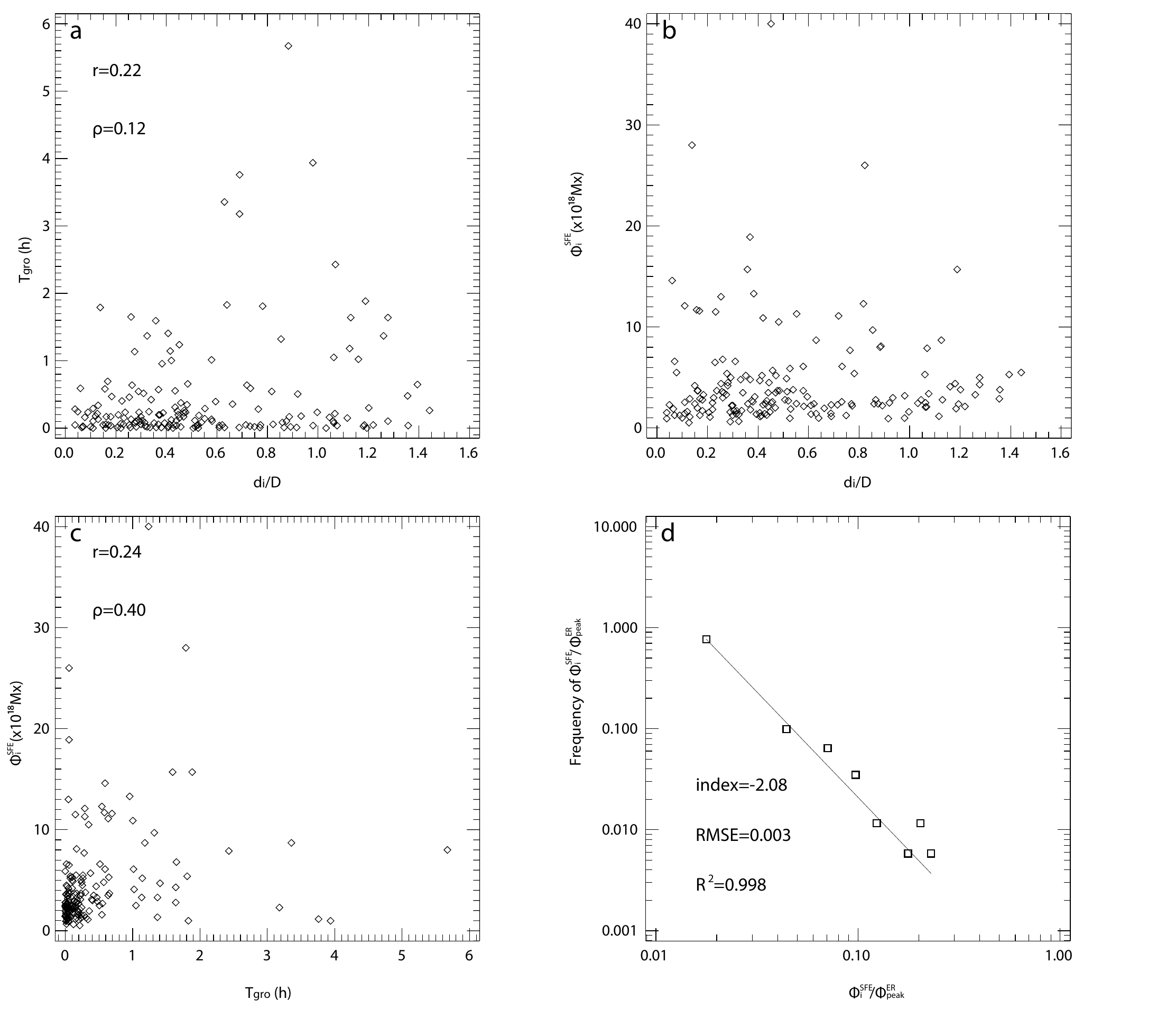}
\caption{Scattered distributions of $\rm{T_{gro}} $—$\rm{d_i}$/$\rm{D}$, $\rm{\Phi^{SFE}_i}$—$\rm{d_i}$/$\rm{D}$, and $\rm{\Phi^{SFE}_i}$—$\rm{T_{gro}} $ in panel a, b, c and logarithmic coordinate frequency distribution of $\rm{\Phi^{SFE}_i}/\rm{\Phi^{ER}_{peak}} $ in panel d  where we employ a bin size of 0.03 . \label{fig:distribution}} 
\end{figure}

Here, we normalize each $\rm{\Phi^{SFE}_i}$ by the $\rm{\Phi^{ER}_{peak}} $ of its corresponding ER and construct their frequency distribution. We find that this distribution approximately conforms to a power law. Therefore, we perform a power-law fitting of the data points which is displayed in panel d of Figure \ref{fig:distribution}. The fitting power-law yields an index of approximately -2.08. To gauge the goodness of fit, we compute the coefficient of determination ($\rm{R^2}$) and the root mean square error (RMSE). The resulting values are 0.998 and 0.003, respectively, indicating that the distribution of the relative flux of SFE, i.e., $\rm{\Phi^{SFE}_i}$/$\rm{\Phi^{ER}_{peak}} $, is likely to conform to a power-law distribution with an index of -2.08 .

\emph{Growth time}: Table \ref{tab:mean value} shows SFEs' growth time ($\rm{T_{gro}} $) from the 5 ERs. Most SFEs grow to their maximum and simultaneously merge into the main polarities and lose their identities within an hour or even several minutes. However, there are also a few SFEs that maintain their independence from the main polarities for a period. Some of them undergo multiple mergers at the same location, while others remain a low magnetic flux level for prolonged periods before undergoing a sudden enhancement, rapidly merging into main polarities. These SFEs may take several hours to reach their maximum flux, and produce the outlier dispersion observed in panels a and c in Figure \ref{fig:distribution }. Furthermore, we calculate the Pearson correlation coefficient r and the Spearman correlation coefficient $\rm{\rho}$ for $\rm{d_i/D-T_{gro}} $ and $\rm{T_{gro}-\Phi^{SFE}_i}$, revealing that in the 172 SFE samples, there is no apparent linear correlation between these parameters . Moreover , a larger growth time of SFEs does not necessarily result in larger amount of magnetic flux. In specific cases, we can observe that certain SFEs can reach flux exceeding $10^{19}$ Mx within minutes, while others may grow for a longer period without significantly higher magnetic flux than most SFEs. This implies that the growth time dies not fundamentally dictate the amount of magnetic flux involved in SFEs. This could be attributed to the inherent flux magnitude of the flux tube branch, determining the strength of SFE. Some branches exhibit fast growth, while others grow more slowly.

\emph{Distance}: Table \ref{tab:mean value} also presents the properties of $\rm{d_i}$. In fact, no SFE with $\rm{d_i}$/$\rm{D}$$>$1 is detected in ER1 and ER2, while in the other ERs, there exist SFEs with 2$>$$\rm{d_i}$/$\rm{D}$$>$1 which could also be confirmed in Figure \ref{fig:distribution}. A larger $\rm{d_i}/\rm{D}$ implies that SFEs are farther away from the center of the main polarities. This may indicate that all SFEs can be observed within a circular area centered on the bipolar's center with a radius of 2$\rm{D}$. This result can be used to delineate the field of view for more occurrences and may serve as a means to verify the validity of ER MHD simulation results. From panel b in Figure \ref{fig:distribution}, it can be observed that most SFEs have a magnetic flux of (1-6)$\times$$10^{18}$ Mx regardless of distance. For those SFEs with stronger magnetic flux, they mainly distribute a distance of $\rm{d_i}/\rm{D}$$<$1. The result perhaps implies that stronger flux tubes are more capable of resisting turbulent convection, allowing them to emerge closer to the main polarities \citep{2012A&A...545A.107B}, and the weaker flux tubes may experience drag and distortion from turbulent convection, causing them to emerge anywhere around ERs \citep{2007A&A...467..703C}.

 \begin{figure}[t]
 \centering
 \includegraphics[width=0.8\textwidth]{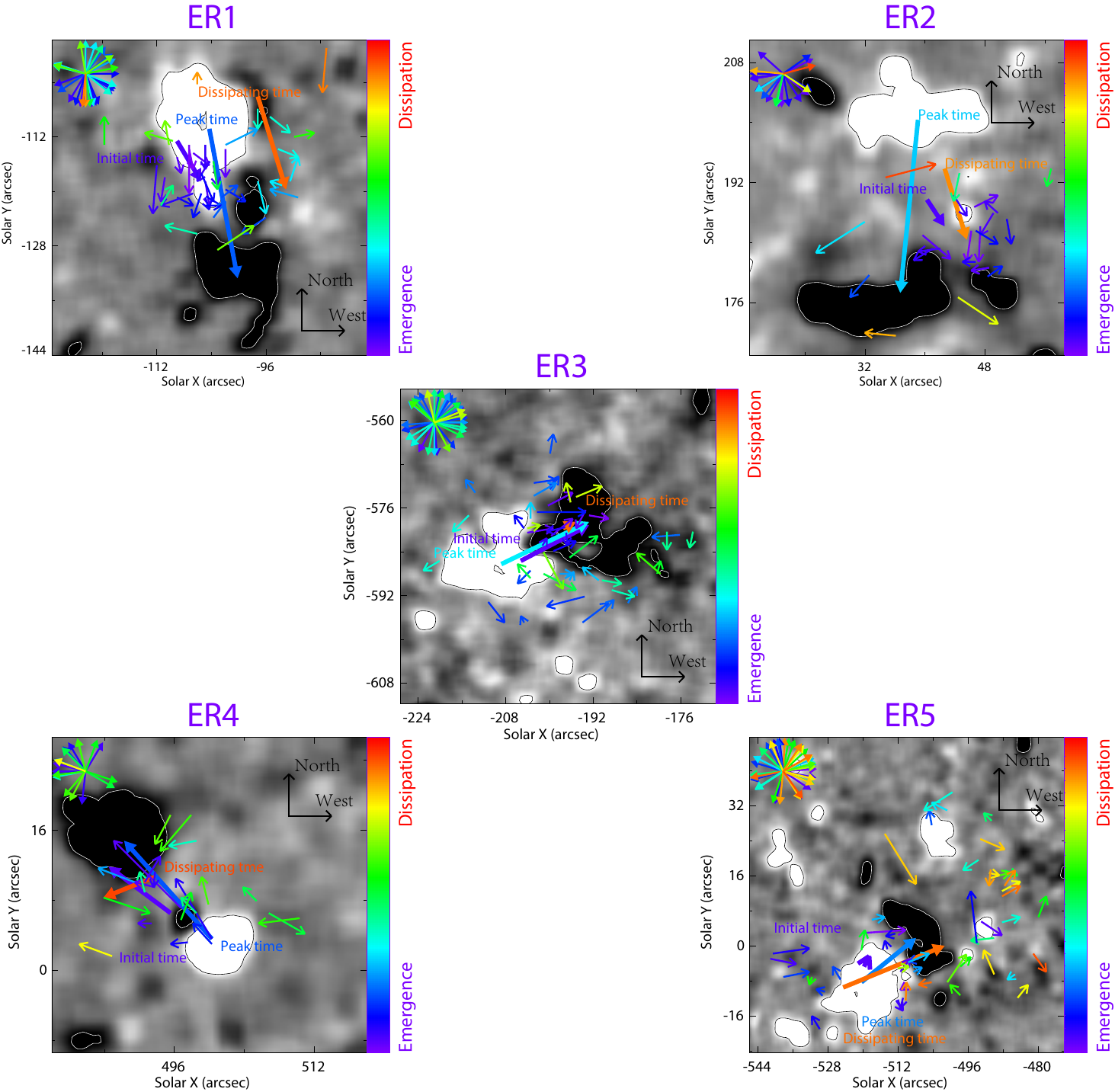}
 \caption{All SFEs with main polarities of ERs marked in magnetograms using arrows. Five magnetograms are selected at the time of maximum magnetic flux for their respective ERs and contour lines with a value of 21 G are used. The arrows start and end at the positive and negative polarities of the SFEs, respectively, and the color indicates the time of SFEs' birth according to the color bar. The main polarities positions during the initial, peak, and dissipation times are also marked with thick arrows.The arrows in the top left corner of each image represent the direction distribution of SFEs within concentric circles.}
 \label{fig:arrow}
\end{figure}

\emph{Orientation}: Figure \ref{fig:arrow} depicts all the magnetic axes of SFEs marked with arrows. The color bar ranging from purple to red represents the chronological sequence of SFEs appearing. The magnetic axis of the ER in initial, peak and dissipation times are marked by thicker arrows in Figure \ref{fig:arrow }. The changes of magnetic axes of ERs in different times seem to imply a slightly twisted structure. It can be found that in the early stage of ER, SFE tends to appear in the middle of the two main polarities. Whereas in the late phase, SFE tends to appear random around ERs.

\begin{figure}[t]
 \centering
 \includegraphics[width=0.9\textwidth]{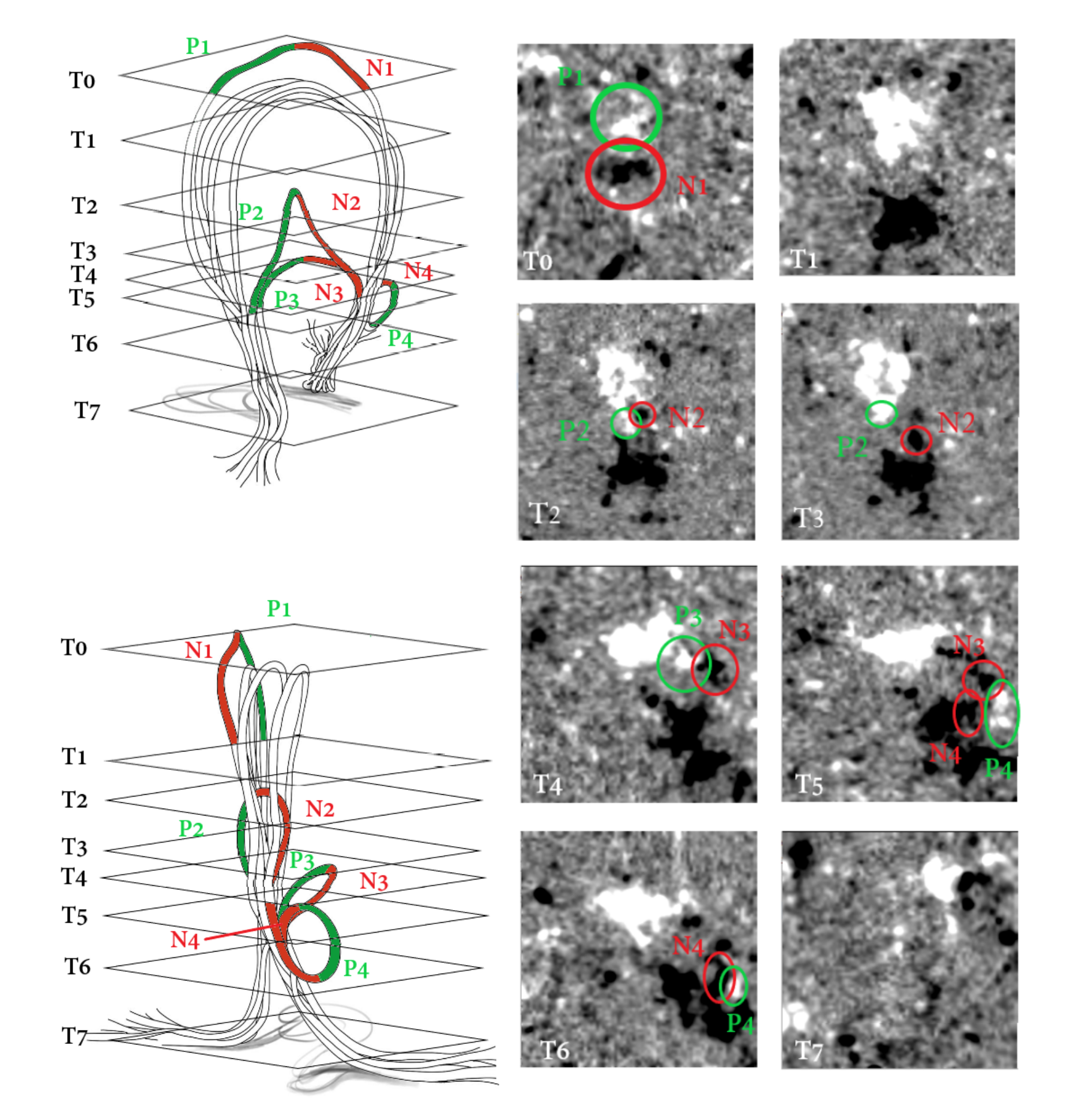}
\caption{Front and side views of magnetic field configurations of the emerging magnetic flux derived from a series of consecutive magnetograms from ER1. The magnetograms spanning from 21-Nov 11:58:30 UT to 21-Nov 21:27:00 UT for T0 to T1, 21-Nov 22:42:45 UT to 22-Nov 01:08:15 UT for T2 to T3, 22-Nov 10:56:15 UT to 22-Nov 14:25:30 UT for T4 to T6 and 24-Nov 00:04:30 UT for T7 are displayed to construct a schematic for subsurface structures. The left part showcases the positive polarities P1, P2, P3, and P4 in green, while the negative polarities are denoted as N1, N2, N3, and N4 in red. The magnetograms on the right part simultaneously label the corresponding positive and negative polarities of each SFE using green and red circles, respectively.}
\label{fig:rope}
\end{figure}

\begin{figure}[t]
 \centering
 \includegraphics[width=0.4\textwidth]{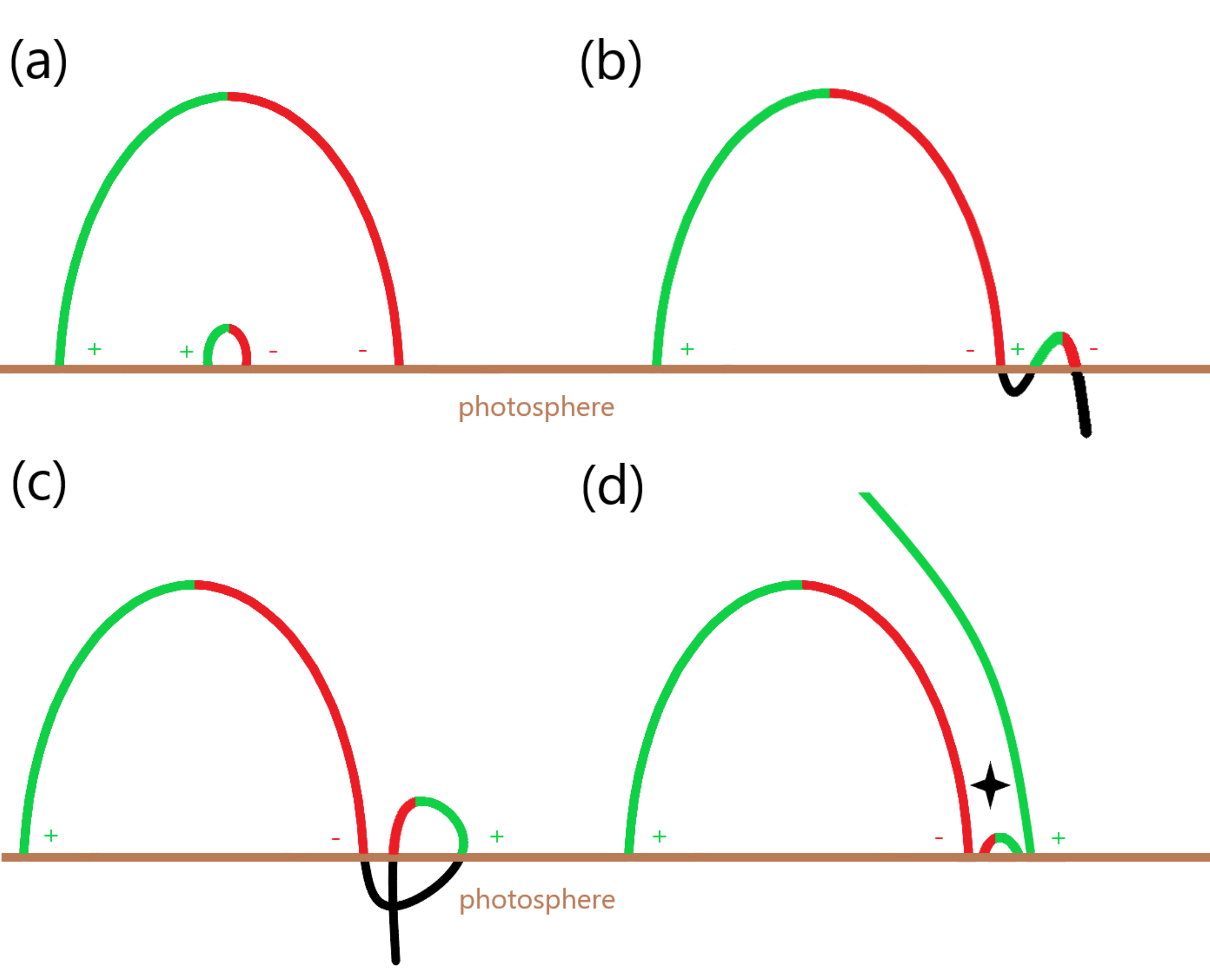}
\caption{Four categories of SFEs. From panel (a) to panel (d) represent regular arch type, kink type, twist type, and bubble type respectively.}
\label{fig:type}
\end{figure}

\section{summary and discussion} \label{sec:4}
This study presents a detailed investigation of SFE process in ERs, based on observations of the line-of-sight magnetic field obtained from SDO/HMI. We select five ERs with the magnetic flux of $10^{20}$ Mx, and analyze their whole evolution. Tens of SFEs are identified in each of ERs by using a visual inspection method. They appear at locations with $\rm{d_i}$/$\rm${D}$<$2, and most of them grow to their maximum magnetic flux within 1 hour. Their magnetic axes exhibit random distribution. The relative flux of SFEs, $\rm{\Phi^{SFE}_i}$/$\rm{\Phi^{ER}_{peak}} $, follows a power law distribution with an index of -2.08, which is similar with the flux distribution at different spatial scales (from intranetwork to active regions) \citep{2009ApJ...698...75P}. This scale-free phenomenon is likely to originate from the near-surface turbulent convection \citep{2011SoPh..269...13T}. Further theoretical and observational validation is required to confirm this hypothesis.

Our observations suggest that the ERs are the concentrative manifestations of chronological emergence for a bunch of magnetic loops below the photosphere, while the SFEs are chronological flux loops, as shown in Figure \ref{fig:rope}. Based on the tracking of the SFEs in the magnetogram, it is speculated that magnetic flux tubes may exhibit not only regular arch structures, but also various types of twisted and distorted shapes (such as P2, N2, P4 and N4 in Figure \ref{fig:rope}). As illustrated in Figure \ref{fig:type}, there are four typical manifestations of SFEs. The first type is the regular arch type, as shown in panel (a). They primarily emerge between main polarities and eventually merge individually with main polarities such as P3 and N3 in Figure \ref{fig:rope}. These SFEs generally exhibit a relatively larger magnetic flux and can even constitute a significant portion of ER. Their magnetic axes have relatively smaller angles with that of ER. The second type is the kink type of SFE, as shown in panel (b) in Figure \ref{fig:type}. It manifests as a bipole where one polarity, situated close to the main polarity, has an opposite sign. Later the polarity with the opposite sign to the main polarity will merge or dissipate, while the other polarity may retain its distinct identity or merge with the main polarity. These SFEs have relatively small magnetic fluxes and small magnetic axis angles with ER. The third type is the twist type as shown in panel (c) in Figure \ref{fig:type}. It manifests as a bipole emergence where one polarity, situated close to the main polarity, has the same sign. As it develops, the polarity of the same sign as the main polarity will merge with the main polarity, while the opposite polarity will cancel with the main polarity or dissipate. These SFEs have small magnetic fluxes, but their magnetic axis angles are generally large, as illustrated by P2, N2 and P4, N4 in Figure \ref{fig:rope}. The fourth type is the bubble type as shown in panel (d) in Figure \ref{fig:type}. They may originate from the magnetic reconnection occurring between ER magnetic fibers and the large-scale overlying magnetic loops. 
They exhibit relatively smaller magnetic flux and are more likely to appear on the outskirts of main polarities, with no particular tendency in magnetic axis angles. However, together with kink type and twist type, they constitute the primary reason for the repetitive counting that leads to the total magnetic flux of SFEs ($\sum\rm{\Phi^{SFE}_i}$) surpassing the overall magnetic flux of the ER ($\rm{\Phi^{ER}_{peak}} $). Considering the limitation of observing only vertical magnetic fields, we assert that it is insufficient to individually verify the category of each SFE. Nevertheless, our confidence in their existence is grounded in the disparity between the sum of magnetic flux of SFEs and the peak magnetic flux of ER, as well as the occurrence of small-scale eruptive activity associated with SFEs. To classify each SFE according to Figure \ref{fig:type}, extreme ultraviolet observations in the lower atmosphere are deemed necessary for a thorough validation of their magnetic connectivity and topology .
As illustrated in Figure \ref{fig:rope}, a bunch of magnetic flux tubes locate below the photosphere, and they are tightly confined by the high-density plasma \citep{2000ApJ...545.1089L}. As it rises, the gas pressure decreases, causing some magnetic flux tube branches to detach from the main tube and emerge independently. As flux tubes arising, they appear on the surface successively as SFEs, which constitute the evolution of ER.

\begin{figure}[t]
 \centering
 \includegraphics[width=0.8\textwidth]{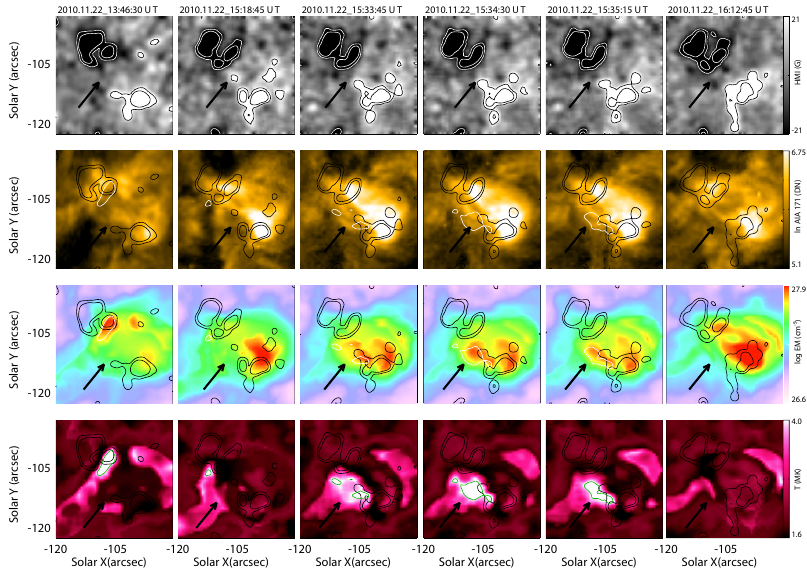}
\caption{An eruption event related to a SFE. The four rows from top to bottom represent the magnetogram, AIA 171 $\rm{\AA}$ image, emission measure (EM), and T respectively, with the calculation methods for EM and T derived from \citet{cheung2015thermal}. In the magnetogram, the positive and negative polarities are marked by white and black contour lines, respectively. The saturation levels are indicated on the far right in the form of color bar. In the AIA 171 $\rm{\AA}$ image channel, the positive and negative polarities in the magnetogram are both depicted using black contours . The magnetic field is outlined in black in the third and fourth rows, with the atmosphere heated beyond 4MK highlighted with white and green lines, respectively.}.
\label{fig:heating}
\end{figure}

The continuous flux emergence in active regions has been extensively studied by various authors. For instance, \citet{1994SoPh..155...99W} and \citet{1996ApJ...462..547L} have elucidated how the flux emergence in active regions is associated with the appearance of electric currents, subsequently leading to solar eruptions. Continuous flux emergence in ERs is also observed, and it is likely to trigger eruptive activity such as jets and ejections of plasmoids. These small-scale explosions could release a significant amount of energy to heat the quiet Sun, making it essential to study magnetic flux evolution in detail within the current spatial resolution limit. In fact, SFEs not only reflect the complex magnetic field configuration of ER, but also are closely related to small-scale activity and atmospheric heating events. As shown in Figure \ref{fig:heating}, an eruption activity directly induced by SFE is observed. The four panels in each column correspond to the magnetic field, AIA 171 $\rm{\AA}$ image, emission measure (EM), and temperature (T), respectively, ordered from left to right according to the time sequence. The EM and T are calculated based on the differential EM inversion method developed by \citet{cheung2015thermal}. It can be seen that at 13:46:30 UT , the position indicated by the arrows do not have positive polarity flux exceeding the threshold flux density of 21 G, and no brightening structure is observed in the same position in the AIA 171 $\rm{\AA}$ image, and the intensities of EM and T are still in a quiet state. Then, a positive polarity patch begins to emerge and reaches its maximum at 15:34:30 UT . An obvious magnetic structure in the closest negative main polarity also appears and simultaneously evolutes with the positive polarity patch. They are identified as a SFE in ER. As the SFE emerges, the domain between the positive and negative polarities begins to intensify in radiation, and the solar atmosphere is gradually heated to over 4MK, which are shown in the images of AIA 171 $\rm{\AA}$, EM and T. During this period, there are also jet-like activity events observed in the AIA 171 $\rm{\AA}$ images, which may be caused by the plasmoid-induced reconnection mentioned in \citet{1995Natur.375...42Y} and \citet{shibata2016fractal} . Subsequently, the positive polarity patch undergoes complete magnetic cancellation, and the radiation and temperature revert to normal levels at 16:12:45 UT . This is an atmospheric heating event caused solely by SFE. This may be attributed to the uplift of a newly emerging loop, which is different from the submergence of the original loop mentioned in \citet{1985ApJ...291..858Z}, \citet{2009ApJ...703.1012Y} and \citet{2012ApJ...752L..24Y}. The uplift leads to magnetic reconnection between the existing loop and the new emerging flux tube. The eruption causes a restructuring of the magnetic field configuration into a lower energy state, accompanied by the release of magnetic energy.

Debate on coronal heating has never ceased. Currently, the most widely accepted heating mechanisms are wave heating and micro/nano-flare heating. During the evolution of an ER, flare-like events occur frequently, which also conform to the observations of microflares or interchange reconnections in \citet{1972SoPh...22...34Z} and \citet{shibata2016fractal}. Next, we will investigate the role of SFEs in heating the solar corona based on statistical analysis.

 SDO is the first mission to be launched for NASA's Living With a Star Program. The authors are grateful to the team members who have made great contributions to the SDO mission.  This work is supported by the National Key R\&D Program of China (grant No. 2022YFF0503000, 2019YFA0405000 and 2021YFA1600500 ), the B-type Strategic Priority Program of the Chinese Academy of Sciences (grant No. XDB41000000), the Key Research Program of Frontier Sciences, CSA (grant No. ZDBS-LY-SLH013), and the National Natural Science Foundation of China (grant No. 12273061 and 12350004).

\bibliography{sample631}{}
\bibliographystyle{aasjournal}

\end{document}